\documentclass[runningheads]{llncs}

\usepackage[T1]{fontenc}
\usepackage{graphicx}
\usepackage{enumitem}
\usepackage{xcolor}
\usepackage[most]{tcolorbox}
\usepackage{tikz}
\usepackage{amsmath} 
\usepackage{hyperref} 
\usepackage{booktabs}
\usepackage{multirow}
\usepackage{algorithm}
\usepackage{algpseudocode}
\usepackage{subfig}

\definecolor{lightgrey}{RGB}{240, 240, 240} 
\definecolor{mygrey}{RGB}{80, 80, 80}     

\tcbset{
  takeaway/.style={
    enhanced,
    colback=lightgrey,
    colframe=lightgrey,
    boxrule=0pt,
    left=1pt,
    right=0pt,
    top=1pt,
    bottom=1pt,
    boxsep=2pt,
    borderline west={1pt}{0pt}{mygrey},
    arc=0mm,
    breakable
  }
}

\usepackage{listings}
\begin{document}

%%
%% Rights management information.
%% CC-BY is default license.
%\copyrightyear{2025}
%\copyrightclause{Copyright for this paper by its authors.
%  Use permitted under Creative Commons License Attribution 4.0
%  International (CC BY 4.0).}

%%
%% This command is for the conference information
%\conference{Joint Proceedings of REFSQ-2025 Workshops, Doctoral Symposium, Posters \& Tools Track, and Education and Training Track. Co-located with REFSQ 2025. Barcelona, Spain, April 7, 2025.}

%%
%% The "title" command
%\title{Hierarchical Feature Clustering for App Review Mining: Enhancing Interpretability and Comparative Analysis with Large Language Models}
%\title{FeClustRE: Hierarchical Clustering and Semantic Tagging of App Features from User Reviews}%

\title{FeClustRE: Hierarchical Clustering and Semantic Tagging of App Features from User Reviews}
\titlerunning{FeClustRE}

\author{Max Tiessler\inst{1}\orcidID{0009-0002-6535-9666} \and Quim Motger\inst{1}\orcidID{0000-0002-4896-7515} }
\institute{
Department of Service and Information System Engineering\\
Universitat Politècnica de Catalunya\\
\email{\{max.tiessler,joaquim.motger\}@upc.edu}}
\maketitle             
% typeset the header of the contribution
% \metatitle*{FeClustRE: Hierarchical Clustering and Semantic Tagging of App Features from User Reviews}
% \metaauthor*{Max Tiessler}
% \metaauthor*{Quim Motger}
% \researchfield*{Requirements Engineering}
% \contribution*{paper class}{technical design paper}
% \contribution*{replication package}{https://doi.org/10.5281/zenodo.17372435}
% \contribution*{code repository}{https://github.com/nlp4se/FeClustRE}
% \contribution*{dataset}{https://github.com/nlp4se/FeClustRE/tree/master/data/input/endpoint_1_process_reviews/ai_assistants}

% Context and motivation
\begin{abstract}
\textbf{[Context and motivation.]} Extracting features from mobile app reviews is increasingly important for multiple requirements engineering (RE) tasks. However, existing methods struggle to turn noisy, ambiguous feedback into interpretable insights. 
\textbf{[Question/problem.]} Syntactic approaches lack semantic depth, while large language models (LLMs) often miss fine-grained features or fail to structure them coherently. In addition, existing methods output flat lists of features without semantic organization, limiting interpretation and comparability.
Consequently, current feature extraction approaches do not provide structured, meaningful representations of app features.
As a result, practitioners face fragmented information that hinder requirement analysis, prioritization, and cross-app comparison, among other use cases.
\textbf{[Principal ideas/results.]} In this context, we propose FeClustRE, a framework integrating hybrid feature extraction, hierarchical clustering with auto-tuning and LLM-based semantic labelling. FeClustRE combines syntactic parsing with LLM enrichment, organizes features into clusters, and automatically generates meaningful taxonomy labels. We evaluate FeClustRE on public benchmarks for extraction correctness and on a sample study of generative AI assistant app reviews for clustering quality, semantic coherence, and interpretability.
\textbf{[Contribution.]} Overall, FeClustRE delivers (1) a hybrid framework for feature extraction and taxonomy generation, (2) an auto-tuning mechanism with a comprehensive evaluation methodology, and (3) open-source and replicable implementation. These contributions bridge user feedback and feature understanding, enabling deeper insights into current and emerging requirements.
\end{abstract}

% % --- SciKGTeX annotations (Open Science Challenge 1) ---
% \iffalse
% \researchproblem{Existing approaches to mining mobile app reviews yield flat, unstructured features; syntactic methods lack semantic depth and LLM-based methods often miss fine-grained, coherent structure, hindering interpretability and comparison.}
% \objective{Develop FeClustRE to extract features, hierarchically cluster them with auto-tuning, and semantically label clusters using LLMs to produce structured, meaningful representations.}
% \method{Hybrid pipeline combining syntactic parsing with LLM enrichment; hierarchical clustering with auto-tuning; automatic taxonomy label generation; evaluation on public benchmarks and a case study of generative AI assistant app reviews.}
% \result{Correctness on public benchmarks and semantically coherent, interpretable clusters observed in the case study.}
% \conclusion{FeClustRE bridges noisy user feedback and feature understanding, supporting requirement analysis, prioritization, and cross-app comparison.}
% \fi
% % --- End SciKGTeX annotations ---

%%
%% Keywords. The author(s) should pick words that accurately describe
%% the work being presented. Separate the keywords with commas.

\begin{keywords}
app stores, feature extraction, hierarchical clustering, large language models, semantic tagging, app reviews, requirements engineering%, taxonomy generation
\end{keywords}

%%
%% This command processes the author and affiliation and title
%% information and builds the first part of the formatted document.

\section{Introduction}
\label{sec:introduction}

Feature extraction (i.e., identifying user-visible functional attributes of a mobile app~\cite{Dabrowski2023}) has gained significant attention over the past decade in the context of app review mining. Requirements engineering (RE) has extensively explored extended use cases derived from analysis of feature mentions, such as %feature-based sentiment analysis~\cite{Dabrowski2020}, %usability analysis~\cite{Bakiu2017}, 
feature prioritization~\cite{MALGAONKAR2022106798}, competition analysis~\cite{Assi2021}, %user rating prediction~\cite{Sarro2018}, 
and software evolution~\cite{Al-Hawari2021}. The state of the art in feature extraction is predominantly represented by syntactic-based approaches~\cite{Dabrowski2023}, which rely on syntactic structures and pattern-matching techniques to extract subsets of tokens within a user review referring to a particular app feature~\cite{Guzman2014,Johann2017,Dragoni2019}. While efficient, replication studies have showcased several limitations in their correctness, such as overlap and ambiguity in syntactic patterns, limited precision and recall, and inability to capture contextual meaning~\cite{Shah2019,Dabrowski2023,Sorathiya2025}.

Meanwhile, large language models (LLMs) offer new opportunities to overcome the limited performance of traditional NLP tasks for RE practices~\cite{Hou2024}. Consequently, several studies have explored leveraging encoder-only LLMs like BERT~\cite{Devlin2018} to improve feature extraction from app reviews~\cite{Wu2021,Araujo2021,Motger2025}. In contrast to syntactic-based challenges, LLM-based feature extraction dynamically adapts to varied linguistic structures and leverages word embeddings that capture semantic nuances, excelling in context-based understanding through transformer architectures. More recently, generative LLMs such as GPT-4 and Llama-2 have also been investigated~\cite{Shah2024}, demonstrating performance comparable to (or in some cases worse than) encoder-only LLMs such as BERT.

\begin{figure*}
    \centering
    \includegraphics[width=0.95\linewidth]{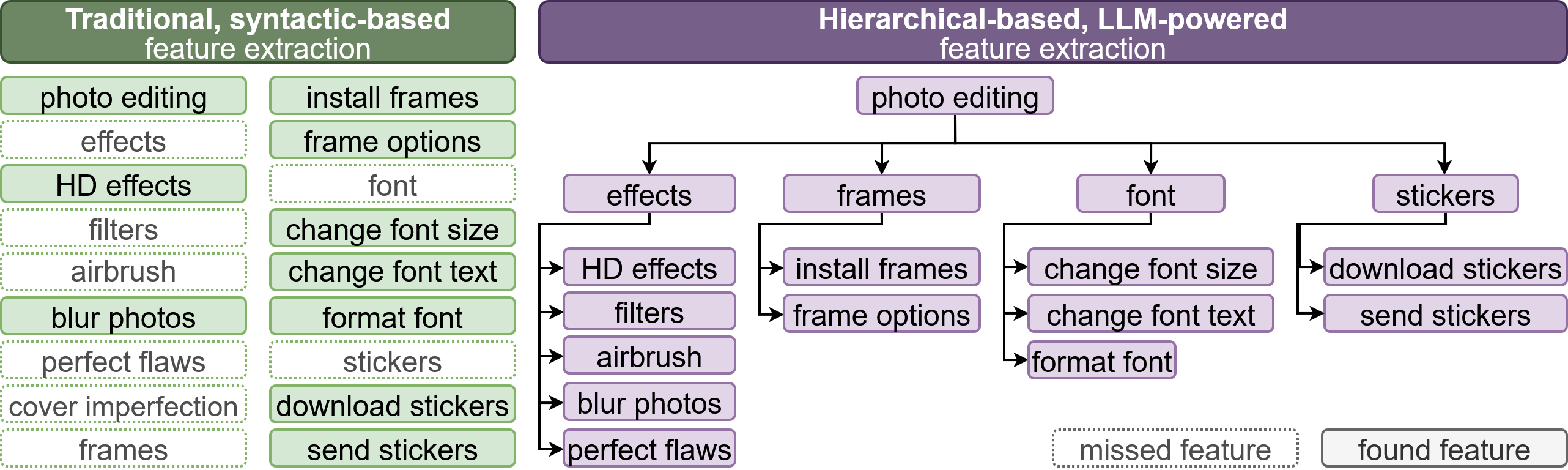}
    \caption{Motivational example of FeClustRE}
    \label{fig:example}
\end{figure*}

In addition to correctness, the adoption of feature extraction tools in practice presents further challenges. Concerning interpretability, previous work output flat lists of features, %~\cite{Guzman2014,Johann2017,DRAGONI20191103,Wu2021,Araujo2021,Gallego2023,Motger2024,Shah2024}
limiting their utility for developers seeking aggregated insights or inter-feature relationships to inform design decisions. This lack of structure is precisely a key limitation. The outputs remain flat and sometimes redundant, hiding semantic relationships between the features. To illustrate this, Figure~\ref{fig:example} shows how traditional methods generate unstructured feature lists, whereas organizing features into a structured, semantically coherent taxonomy with meaningful labels makes relationships clearer and easier to interpret.

To bridge this gap, we present \textit{FeClustRE}, a comprehensive framework for feature clustering and semantic tagging of mobile app review features via hierarchical taxonomies. Our framework addresses this gap through an end-to-end solution, integrating multiple components into a unified workflow. Our contribution is three-fold. First, we design and implement a feature extraction pipeline that integrates syntactic and LLM-based state-of-the-art methods. Second, we develop an auto-tuning hierarchical clustering framework that automatically identifies optimal clustering parameters, creates semantically coherent taxonomies through automatic merging mechanisms, and generates meaningful cluster labels with an LLM-based semantic tagging. Third, we provide an open-source implementation and evaluation that demonstrate its potential applicability to real-world RE scenarios. Finally, our evaluation combines public datasets to benchmark the hybrid feature extraction correctness and a sample study on generative AI assistant app reviews to assess clustering quality, semantic coherence, and interpretability. All materials derived from this research are openly shared (see \textit{Data Availability Statement} at the end of this paper).

% --- SciKGTeX Introduction Annotations ---
%\researchproblem{Existing feature extraction approaches from app reviews output flat, sometimes redundant lists of features, limiting interpretability and obscuring semantic relationships.}
%\researchproblem{Syntactic methods suffer from ambiguity and low recall, while LLM-based methods, though more adaptive, still lack structured outputs.}

%\objective{Bridge this gap with FeClustRE, a comprehensive framework for feature clustering and semantic tagging of app review features via hierarchical taxonomies.}

%\method{Design an end-to-end pipeline that integrates syntactic and LLM-based extraction methods, auto-tuned hierarchical clustering, and LLM-based semantic tagging for taxonomy generation.}
% --- End SciKGTeX Introduction Annotations ---

\section{Related Work}
\label{sec:background}

% Syntactic-based Feature Extraction
Syntactic-based feature extraction methods rely on linguistic patterns such as part-of-speech (POS) structures~\cite{Guzman2014,Johann2017,Dragoni2019,Gallego2023}. However, replication studies reveal several performance limitations~\cite{Shah2019,Dabrowski2023}. These methods are highly sensitive to dataset-specific syntactic patterns and often require extensive customization for different app categories, which limits their adaptability across domains~\cite{Dabrowski2023}. They also lack the ability to capture semantic and contextual knowledge~\cite{Shah2019}, leading to many false positives (e.g., \textit{nice app}) and missed feature mentions (e.g., \textit{synchronization})~\cite{Dabrowski2023}. Ambiguous or overlapping POS patterns often yield redundant or inaccurate extractions (e.g., \textit{download stuff})~\cite{Shah2019}. Performance also degrades in noisy text with typos, abbreviations, or slang (e.g., \textit{focus functn})~\cite{Johann2017}. Finally, syntactic methods frequently miss single-word (e.g., \textit{note-taking}) and multi-noun features (e.g., \textit{in-app purchases}), lowering overall recall~\cite{Dabrowski2023}.

% LLM-based Feature Extraction
To address some of these issues, solutions have turned to encoder-only LLMs, leveraging contextual embeddings to identify features based on semantic understanding~\cite{Wu2021,Araujo2021,Motger2025}. While they overcome certain syntactic limitations, they introduce new challenges. LLM-based methods typically depend on domain-specific fine-tuning, as generic models underperform when applied to specialized app categories without adaptation~\cite{Araujo2021}. Their performance also degrades on noisy data, where abbreviations or misspellings interfere with embeddings~\cite{Wu2021}. Moreover, they struggle with fine-grained features (e.g., \textit{take off unwanted bits captured in the photo}), leading to low precision under exact or partial matching~\cite{Motger2025}. Rare or emerging features (e.g., \textit{GPX track}) are also difficult to capture because they appear infrequently in training data~\cite{Wu2021}. Finally, encoder-only models often favor frequent patterns, biasing results toward common features while overlooking less frequent but important ones~\cite{Motger2025}.

% Feature Organization and Practical Application Gaps
In addition, both syntactic and LLM-based approaches share limitations. They generally produce flat feature lists without meaningful organization. This makes it difficult for developers to identify user needs or to understand the feature set of an application as a whole. While some approaches either apply~\cite{Guzman2014} or envision~\cite{Khubaib2024} topic modelling, these efforts are limited to grouping fine-grained features without providing semantic structure or hierarchical relationships. More recently, Jin et al. introduced a framework that automatically constructs multi-level feature trees from software artifact libraries by combining clustering algorithms with LLM-based summarization~\cite{Jin2025FTBuilder}. Unlike prior work, their contribution builds hierarchical structures, providing semantic organization and demonstrating improvements in artifact recommendation. However, its evaluation is centred on software artifacts rather than user-generated reviews, where input is noisier, domain-specific, and highly informal. %In such settings, developers often struggle to identify key themes, prioritize issues, or understand feature dependencies~\cite{Dabrowski2023}, which limits the utility of feature extraction for producing aggregated insights or feature relationships to inform design decisions~\cite{Dabrowski2022}.

Despite advances, the practical application of feature extraction remains limited. Most methods are still evaluated in controlled settings~\cite{Guzman2014,Johann2017,Dragoni2019,Wu2021,Araujo2021,Gallego2023,Motger2025,Shah2024}, with little focus on real-world use cases that turn extracted features into actionable insights. This gap is most evident beyond feature-based sentiment analysis, which remains the dominant application~\cite{Dabrowski2023}. Promising alternatives such as app benchmarking~\cite{AlvesdeLima2022} and competition analysis~\cite{Dalpiaz2019} %, and market trend identification~\cite{Motger2024a} 
require structured, interpretable feature insights. Yet current tools lack integrated visualization, cross-application comparison, and scalable architectures for large-scale data. Without frameworks that combine extraction, semantic organization, and interactive exploration, adoption remains limited, as practitioners need comprehensive solutions rather than isolated prototypes.

% --- SciKGTeX Related Work Annotations ---
%\researchproblem{Syntactic-based feature extraction methods are highly sensitive to dataset-specific linguistic patterns, lack semantic context, and degrade on noisy or informal text.}
%\researchproblem{LLM-based feature extraction approaches require domain-specific fine-tuning and struggle with rare or fine-grained features, leading to incomplete extraction.}
%\researchproblem{Both syntactic and LLM-based methods typically output flat, unstructured feature lists, limiting interpretability and practical adoption in requirements engineering.}
%\researchproblem{Existing frameworks rarely evaluate on real-world, practitioner-oriented datasets, reducing the transferability and impact of their findings.}
% --- End SciKGTeX Related Work Annotations ---

\section{System Design \& Architecture}
\label{sec:design}

\subsection{Research Goal}

The goal of \textit{FeClustRE} is to advance the state-of-the-art in feature extraction from app reviews by addressing two aspects: (1) improving the \textbf{correctness} of feature extraction methods, and (2) enhancing the \textbf{structuring} and \textbf{interpretability} of extracted features.
To achieve these goals, and based on the challenges identified in Section \ref{sec:background}, we propose the following research questions:

\begin{enumerate}[label=\textbf{RQ\textsubscript{\arabic*}.}]
    \item Does combining syntactic and LLM-based methods improve feature extraction correctness in app reviews?
    \item How does hierarchical clustering help organize and interpret features extracted from app reviews?
\end{enumerate}

To address \textbf{RQ\textsubscript{1}}, we built on previous work to design a hybrid pipeline integrating a syntactic-based method and an LLM-based feature extraction method under a common pre-processing stage. 
Given that feature extraction is a well-established task with publicly available benchmarks, we conducted a quantitative analysis on two annotated app review datasets. %Since recall is often prioritized over precision in this RE task~\cite{Berry2021}, we report results using the balanced F-measure proposed in prior work~\cite{Motger2025}. 
The specific evaluation settings and baselines are detailed in Section~\ref{sec:evaluation}.

To address \textbf{RQ\textsubscript{2}}, we designed and evaluated the performance of an auto-tuning hierarchical clustering framework combined with LLM semantic tagging over extracted features that organized in interpretable taxonomies. This evaluation follows a two-fold strategy. First, we conducted a quantitative evaluation over our auto-tuning approach with clustering internal quality metrics~\cite{heumann2024taxonomyclustering}. Second, we qualitatively evaluated the semantic quality of LLM-generated taxonomy labels through distinctiveness analysis, similarity-based merging effectiveness, and coherence evaluation \cite{Unterkalmsteiner2023Taxonomy}.

\subsection{System Architecture}
\begin{figure*}[h]
    \centering
    \includegraphics[width=\linewidth]{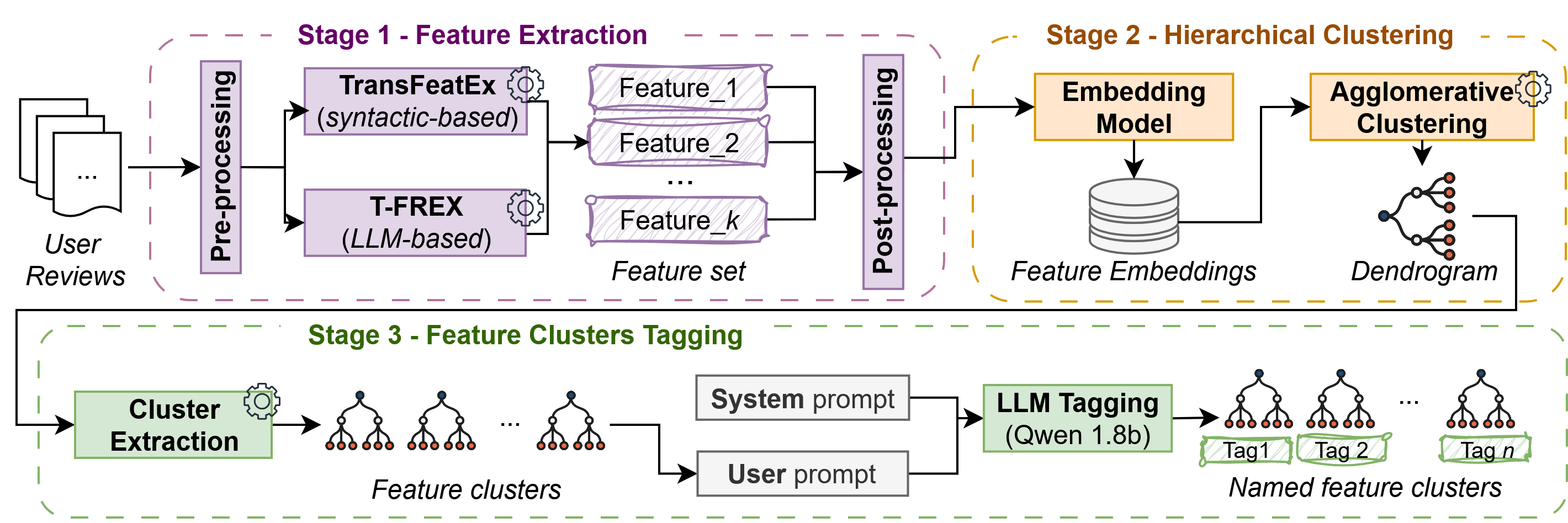}
    \caption{FeClustRE three-stage pipeline.}
    \label{fig:system-design}
\end{figure*}

In order to explore both research questions, we designed a three-stage pipeline within the FeClustRE framework. Figure~\ref{fig:system-design} illustrates the pipeline design.

\subsubsection{Stage 1 - Feature Extraction.}  
Given a review set $\mathcal{R}$ as input, we extract features from the reviews using an ensemble of feature extraction methods, integrated as standalone microservices. Based on the context of this research, we apply a hybrid approach that combines syntactic and LLM-based extraction, using two representative methods from our previous work: (1) for the syntactic method, we use TransFeatEx~\cite{Gallego2023}, an open-source tool that applies syntactic strategies~\cite{Dabrowski2023} enhanced with encoder-only LLMs to provide linguistic annotations such as POS tags, and (2) for the LLM-based method, we use T-FREX~\cite{Motger2025}, a fine-tuned encoder-only model for token classification. Specifically, we employ the BERT-base version of T-FREX for its balance between correctness and efficiency.  
Before feature extraction, reviews are pre-processed by removing emojis, filtering URLs, and applying normalization, while keeping punctuation for feature boundary detection. After extraction, we apply a post-processing step to normalize and deduplicate features.  

The processed review set $\mathcal{R}$ is finally transformed into a feature set $\mathcal{F} = \{f_1, f_2, \dots, f_n\}$.  

\subsubsection{Stage 2 - Hierarchical Clustering.}
\begin{algorithm}[t]
\caption{Hierarchical Clustering with Auto-Tuning}
\label{alg:dendrogram-generation}
\begin{algorithmic}[1]
\Require 
   \Statex Feature set $\mathcal{F} = \{f_1, f_2, \dots, f_n\}$, Embedding function $e$, Distance metric $d$, Linkage method $\lambda$  
\Ensure 
   \Statex Clustering candidates $\mathcal{C}$ with quality metrics
\vspace{0.25em}
   \State \textbf{(1) Compute Embeddings}: $\mathcal{E} \gets e(\mathcal{F})$    
   \State \textbf{(2) Compute Affinity Matrix}: $\mathbf{A} \gets \text{Affinity}(\mathcal{E}, d)$
   \State \textbf{(3) Generate Linkage Matrix} $\mathbf{L} \gets \text{Linkage}(\mathbf{A}, \lambda)$  
   \State \textbf{(4) Construct dendrogram} $\mathcal{D} \gets \text{Dendrogram}(\mathbf{L})$
   \State \textbf{(5) Auto-tune clustering}:
   \For{threshold $t$ in $[0.1, 0.9]$}
       \State $clusters \gets \text{Cut}(\mathcal{D}, t)$
       \State $silhouette \gets \text{SilhouetteScore}(\mathcal{E}, clusters)$
       \State $davies\_bouldin \gets \text{DaviesBouldinScore}(\mathcal{E}, clusters)$
       \State $score \gets \text{CompositeScore}(silhouette, davies\_bouldin, |clusters|)$
       \State $\mathcal{C} \gets \mathcal{C} \cup \{(clusters, score, metrics)\}$
   \EndFor
   \State \Return $\mathcal{C}$
\end{algorithmic}
\end{algorithm}

The second stage applies Algorithm~\ref{alg:dendrogram-generation}, which describes the hierarchical agglomerative clustering process carried out for the set of features $\mathcal{F}$ obtained in the previous stage. 
First, the features in $\mathcal{F}$ are embedded in a high-dimensional space using using configurable embedding models (e.g., Sentence-BERT~\cite{Reimers2019}, Sentence-T5~\cite{ni2021sentencet5scalablesentenceencoders}) to generate semantic embeddings $\mathcal{E}$.
Second, an affinity matrix $\mathbf{A}$ is computed using cosine dissimilarity metric $d$, which measures the semantic distance between each pair of features. 
Third, average linkage method $\lambda$ is applied to $\mathbf{A}$, producing a linkage matrix $\mathbf{L}$ that encodes the hierarchical structure of $\mathcal{F}$. 
Fourth, $\mathbf{L}$ is used to construct a complete dendrogram $\mathcal{D}$ of the set of features $\mathcal{F}$. The system then applies an auto-tuning mechanism that evaluates multiple height thresholds to cut $\mathcal{D}$ into clustering candidates, assessing each configuration using silhouette score and Davies-Bouldin index \cite{heumann2024taxonomyclustering} to identify optimal cluster arrangements through composite scoring. Higher silhouette scores indicate denser, well-separated clusters, while lower Davies-Bouldin index values reflect more cohesive and distinct feature groupings.
The result of this stage is a set of clustering candidates $\mathcal{C}$.%, each associated with quality metrics for evaluation in the third stage.

\subsubsection{Stage 3 - Feature Clusters Tagging.}
In the third stage, the framework automatically selects the most suitable clustering configuration from the candidate set $\mathcal{C}$ generated in the previous stage.
To accommodate different usage scenarios, \textit{FeClustRE} provides three strategies that users can select from:
(1) the \textit{silhouette} strategy prioritizes internal cluster coherence by selecting candidates with the highest silhouette score; 
(2) the \textit{balanced} strategy combines multiple quality metrics, including silhouette score, inverted Davies-Bouldin score, and penalties for cluster count and cluster size; and (3) the \textit{conservative} strategy favors stability by promoting fewer, larger clusters with strong cohesion. 
%These strategies are designed to be adaptable and can be extended to other application domains. They 
%eliminate the need for domain-specific knowledge about the apps being clustered, allowing the framework to produce predictable results without relying on manual tuning. 
Once a clustering configuration is selected, the pipeline applies LLM semantic tagging and taxonomy merging to create feature taxonomies, as described in Algorithm~\ref{alg:cluster-extraction}.

\begin{algorithm}[t]
\caption{LLM Semantic Tagging and Taxonomy Merging}
\label{alg:cluster-extraction}
\begin{algorithmic}[1]
\Require Selected clustering $\mathcal{C} = \{C_1, C_2, \dots, C_k\}$, 
feature set $\mathcal{F}$, large language model $g$, 
similarity threshold $\sigma$
\Ensure Consolidated taxonomies $\mathcal{T} = \{T_1, T_2, \dots, T_m\}$
\vspace{0.25em}

\State \textbf{(1) Semantic Labeling:}
\For{each cluster $C_j \in \mathcal{C}$}
   \State Retrieve feature subset $\mathcal{F}_j \subset \mathcal{F}$ assigned to $C_j$
   \State Generate descriptive label $l_j \gets g(\mathcal{F}_j)$ using few-shot prompting
   \State Create mini-taxonomy: $T_j \gets \text{BuildHierarchy}(\mathcal{F}_j, l_j)$
\EndFor

\State \textbf{(2) Taxonomy Merging:}
\For{each pair $(T_i, T_j)$ of mini-taxonomies}
   \State $similarity \gets \text{CosineSimilarity}(\text{embed}(l_i), \text{embed}(l_j))$
   \If{$similarity \geq \sigma$}
       \State $T_{\text{merged}} \gets \text{MergeTaxonomies}(T_i, T_j)$
       \State Remove $T_i, T_j$ and add $T_{\text{merged}}$ to $\mathcal{T}$
   \EndIf
\EndFor
\State \textbf{return} $\mathcal{T}$
\end{algorithmic}
\end{algorithm}

First, the selected clustering configuration provides a set of clusters 
$\mathcal{C} = \{C_1, C_2, \dots, C_k\}$, where each cluster $C_j$ contains 
semantically related features from the feature set $\mathcal{F}$.
Second, for each cluster $C_j$, the corresponding feature subset 
$\mathcal{F}_j$ is passed to a generative LLM $g$ (e.g., Qwen 1.8b), which generates 
a descriptive label $l_j$ representing the semantic category of the cluster 
using few-shot prompting\footnote{Prompt template and examples available in the replication package.} with domain-specific examples from related apps (see Section~\ref{sec:datasets}).
Third, the labelled clusters are converted into small taxonomies by building 
hierarchical tree structures that preserve internal feature relationships. 
Fourth, a merging mechanism evaluates semantic similarity between taxonomies, considering both structural alignment and label embeddings. Taxonomies that exceed a similarity threshold $\sigma$, which can be defined by domain experts, are automatically merged, with the larger taxonomy serving as the primary structure and smaller ones integrated as
sub-branches.

The resulting output is a set of taxonomies 
$\mathcal{T} = \{T_1, T_2, \dots, T_m\}$ with semantic labels and hierarchical  relationships, stored in a graph database for efficient querying and visualization.

\section{Evaluation Design}
\label{sec:evaluation}

\subsection{Experimental Setting}
To address \textbf{RQ\textsubscript{1}}, we perform a controlled comparison to validate the effectiveness of our hybrid feature extraction approach across three configurations: (1) a syntactic-based baseline using TransFeatEx, (2) an LLM-based baseline using T-FREX, and (3) our hybrid approach integrating both methods. Each method is evaluated against the same feature extraction benchmarks (see Section~\ref{sec:datasets}) using a common preprocessing pipeline including text cleaning, normalization, and tokenization. Performance was measured through precision, recall, and F-score calculated using the weighting factor $\beta = 2.385$, as proposed for feature extraction tasks in previous work~\cite{Motger2025}.

To address \textbf{RQ\textsubscript{2}}, we evaluate the full auto-tuning hierarchical clustering framework with LLM semantic tagging through a two-step process that involves: (1) quantitative analysis of clustering quality using internal metrics (e.g., silhouette score, Davies-Bouldin index) together with an evaluation of the auto-tuning effectiveness; and (2) semantic qualitative assessment of LLM-generated taxonomy labels and hierarchical structure.

\subsection{Datasets}
\label{sec:datasets}

For \textbf{RQ\textsubscript{1}} we established feature extraction benchmarks from mobile app reviews to ensure a robust validation of our hybrid approach. The first dataset (\textit{expert}) contains 2,062 reviews~\cite{Dabrowski2023}, annotated by developers through multiple iterations to establish a reliable ground truth for feature extraction.The second dataset (\textit{crowdsourced}) consists of 27,780 app reviews~\cite{Motger2025}, annotated via a label transfer mechanism that leverages crowdsourced feature annotations from real users in a mobile app recommender platform.

For \textbf{RQ\textsubscript{2}}, we conducted a sample study evaluation on \textit{generative LLM-based chatbot} mobile apps to evaluate our framework on a contemporary, evolving domain. Generative LLM-based chatbots (e.g., ChatGPT) leverage LLMs to provide conversational support, task automation, and information retrieval~\cite{RAY2023121}. %These systems handle multi-domain input, generate meaningful responses, and improve through adaptive learning~\cite{DWIVEDI2023102642}. 
%Within the mobile ecosystem, LLM-powered assistants are emerging as key productivity tools with cross-domain impact~\cite{Kimberly2025}. Moreover, their applications are actively explored in %education~\cite{Sampaio2024}, healthcare~\cite{fi15090286}, 
%software engineering~\cite{Hou2024}, and RE~\cite{Arora2024}. 
The selection of mobile apps was guided by four criteria: (\textit{i}) popularity, ensuring substantial review volumes~\cite{Hou2024}; (\textit{ii}) disruptive market impact, as they redefine assistant capabilities; (\textit{iii}) relevance for feature discovery, through frequent introduction of novel functionalities; and (\textit{iv}) potential for market analysis, as this segment reflects emerging trends and user demands. Based on these criteria, we selected seven representative apps and collected their user reviews from Google Play over a one-month span: %OpenAI's ChatGPT, Claude by Anthropic, DeepSeek AI Assistant, Google's Gemini, Le Chat by Mistral, Microsoft's Copilot, and Perplexity Ask Anything. In total, we collected 158,207 reviews: 
OpenAI's ChatGPT (119,892 reviews), Google's Gemini (26,883), Microsoft's Copilot (5,323), Perplexity by Sonar (3,818), DeepSeek AI Assistant (1,334), Claude by Anthropic (882), and Le Chat by Mistral (75). Reviews were gathered on July 17th, 2025 using an app review collection service integrating multiple mobile sources~\cite{Motger2024b}.  

To evaluate the effect of data quantity on the quality of the framework's results, and to validate that both small and large samples can produce reliable results while also assessing how quality improves as more data are fed into the pipeline, we adopted two sampling strategies: (1) subsets of 2K and (2) 50K reviews. These subsets were generated at runtime from the full dataset using stratified sampling to preserve app-level proportions.

Finally, to perform semantic taxonomy tagging without relying on expert annotation, we used a few-shot prompting approach based on features extracted from official developer descriptions. These descriptions, defined by the app developers themselves, serve as a close approximation to a gold standard. The features were manually extracted from the application descriptions to build training examples for the few-shot prompts, and the validation was manually reviewed independently by both authors. The prompts included examples from multiple apps within the same market segment to guide the model in generating contextually appropriate and semantically consistent category names.

\section{Results}
\label{sec:results}

\subsection{Feature Extraction Correctness (RQ\textsubscript{1})}

Table~\ref{tab:correctness} reports the correctness metrics for our hybrid feature extraction approach compared to syntactic-based (TransFeatEx) and LLM-based (T-FREX) baselines. We used the feature matching method proposed by Dabrowski et al.~\cite{Dabrowski2023}, which considers features as matching if one is identical to or a subset of the other, with an allowed length difference of up to $n$ words: $n=0$ for exact matches, $n=1$ for minor differences, and $n=2$ for slightly larger variations.

\begin{table*}[h]
\small
\caption{Correctness results for feature extraction}
\begin{tabular}{@{}llccccccccc@{}}
\toprule
\textbf{Dataset} & \textbf{Method}
  & \multicolumn{3}{c}{\textbf{Exact (n=0)}} &
 \multicolumn{3}{c}{\textbf{Partial 1 (n=1)}} &
 \multicolumn{3}{c}{\textbf{Partial 2 (n=2)}} \\
\textbf{} &
 \textbf{} &
 \textbf{P} &
 \textbf{R} &
 \textbf{F$_\beta$} &
 \textbf{P} &
 \textbf{R} &
 \textbf{F$_\beta$} &
 \textbf{P} &
 \textbf{R} &
 \textbf{F$_\beta$} \\ \midrule
\multirow{3}{*}{\textbf{Expert}} 
& \textbf{Syntactic} &  0.027 & 0.022 & 0.023 & 0.197 & 0.165 & 0.169 & 0.248 & 0.208 & 0.213 \\
& \textbf{LLM}       &  \textbf{0.211} & 0.078 & 0.086 & \textbf{0.386} & 0.142 & 0.157 & \textbf{0.471} & 0.174 & 0.192 \\
& \textbf{Hybrid}    &   0.082 & \textbf{0.099} & \textbf{0.096} & 0.238 & \textbf{0.286} & \textbf{0.278} & 0.291 & \textbf{0.348} & \textbf{0.338} \\
\midrule
\multirow{3}{*}{\textbf{Crowd.}}    
& \textbf{Syntactic} &  0.022 & 0.021 & 0.021 & 0.162 & 0.140 & 0.143 & 0.263 & 0.226 & 0.231 \\
& \textbf{LLM}       &  \textbf{0.722} & 0.661 & 0.669 & \textbf{0.739} & 0.676 & 0.685 & \textbf{0.739} & 0.677 & 0.685 \\
& \textbf{Hybrid}    &  0.472 & \textbf{0.746} & \textbf{0.686} & 0.489 & \textbf{0.774} & \textbf{0.712} & 0.496 & \textbf{0.787} & \textbf{0.724} \\ 
\midrule
\multirow{3}{*}{\textbf{Avg.}} 
& \textbf{Syntactic} &  0.025 & 0.022 & 0.022 & 0.180 & 0.153 & 0.156 & 0.256 & 0.217 & 0.222 \\
& \textbf{LLM}       &  \textbf{0.467} & 0.370 & 0.378 & \textbf{0.563} & 0.409 & 0.421 & \textbf{0.605} & 0.426 & 0.439 \\
& \textbf{Hybrid}    &  0.277 & \textbf{0.423} & \textbf{0.391} & 0.364 & \textbf{0.530} & \textbf{0.495} & 0.394 & \textbf{0.568} & \textbf{0.531} \\ 
\bottomrule
\end{tabular}
\centering
\label{tab:correctness}
\end{table*}

Results demonstrate that our hybrid method consistently achieves the highest recall across all datasets and evaluation settings, confirming the hypothesis that combining syntactic and LLM-based approaches significantly reduces missed features (false negatives). While the LLM-based method in isolation achieves higher precision, the hybrid setting achieves the highest balanced F-measure across all settings.
For partial matches, which represent practical feature extraction scenarios, the hybrid method achieves the best F-score performance on both datasets, leading to an average F-score of 0.495 (n=1) and 0.531 (n=2).

\begin{tcolorbox}[takeaway]
\textbf{Takeaway 1.} Hybrid feature extraction combining syntactic and LLM-based methods can potentially improve feature correctness, particularly by reducing the amount of false negatives.
\end{tcolorbox}

\subsection{Feature Clustering Interpretability (RQ\textsubscript{2})}

Table~\ref{tab:rq2_summary} summarizes clustering quality and granularity for LLM-based chatbot apps across different feature extraction models and embeddings.
The results reveal three complementary perspectives. TransfeatEx achieves the highest cohesion/separation (silhouette up to 0.28) but produces coarser taxonomies with fewer clusters (18 to 24 per application), making it suitable for high-level analysis. T-FREX strikes a balance, with larger numbers of clusters (60 to 227) and stable cohesion. The hybrid setting, while slightly below TransfeatEx in cohesion (silhouette $\sim$0.19), generate much richer taxonomies (80--301 clusters), making them particularly effective for practitioners needing fine-grained feature understanding. This aligns with the insights from RQ\textsubscript{1}, where the hybrid approach offered the most balanced performance, achieving high recall and strong overall F-measure while maintaining competitive precision.

\begin{table*}[t]
\caption{Clustering performance at 2K vs. 50K samples}
\centering
\begin{tabular}{@{}llcccc@{}}
\toprule
\textbf{Model} & \textbf{Embedding} 
& \textbf{Sample} 
& \textbf{Silhouette} 
& \textbf{Davies-B} 
& \textbf{Clusters} \\ 
\midrule
Hybrid & AllMiniLM   & 2K  & 0.19 $\pm$ 0.02 & 0.96 & 88  \\
       &             & 50K & 0.19 $\pm$ 0.03 & 0.79 & 273 \\
Hybrid & Sentence-T5 & 2K  & 0.18 $\pm$ 0.01 & 1.01 & 80  \\
       &             & 50K & 0.19 $\pm$ 0.03 & 0.76 & 301 \\
\midrule
T-FREX & AllMiniLM   & 2K  & 0.18 $\pm$ 0.04 & 0.91 & 69  \\
       &             & 50K & 0.19 $\pm$ 0.05 & 0.87 & 227 \\
T-FREX & Sentence-T5 & 2K  & 0.18 $\pm$ 0.04 & 0.92 & 60  \\
       &             & 50K & 0.19 $\pm$ 0.05 & 0.87 & 227 \\
\midrule
TransfeatEx & AllMiniLM   & 2K  & 0.20 $\pm$ 0.04 & 0.96 & 19  \\
            &             & 50K & 0.28 $\pm$ 0.14 & 0.82 & 24 \\
TransfeatEx & Sentence-T5 & 2K  & 0.19 $\pm$ 0.04 & 0.96 & 19  \\
            &             & 50K & 0.28 $\pm$ 0.14 & 0.82 & 24 \\
\bottomrule
\end{tabular}
\label{tab:rq2_summary}
\end{table*}

On the other hand, the auto-tuning mechanism is particularly valuable in eliminating manual parameter selection while maintaining clustering quality across different app categories and dataset sizes. In our experiments, it generated between 80 and 301 clusters while keeping silhouette scores stable (around 0.19). This indicates that the framework is robust to varying feature densities and semantic complexities across different applications.

\begin{tcolorbox}[takeaway]
\textbf{Takeaway 2.} Hierarchical clustering with hybrid feature extraction enabled with auto-tuning settings allows different feature-level clustering granularities, facilitating domain and use case adaptation.
\end{tcolorbox}

Concerning data quantity, TransfeatEx strengthens cohesion with more data (silhouette 0.19 to 0.28) while maintaining compact taxonomies (19 to 24 clusters). T-FREX maintains stable cohesion but substantially increases granularity (69 to 227 clusters). Hybrid extraction show the largest growth in detail while holding cohesion steady (around 0.19), expanding from 80 to more than 270 clusters. This progression shows that scaling data improves both cohesion and granularity. However, results with a small subset of reviews (2K) produce comparable internal cohesion and cluster distinctiveness scores, showcasing that FeClustRE is also applicable in contexts with a small number of reviews.

\begin{tcolorbox}[takeaway]
\textbf{Takeaway 3.} Larger review datasets improve internal cluster cohesion and external distinctiveness. However, FeClustRE can still achieve comparable results with limited review data.
\end{tcolorbox}

From a structural perspective, the taxonomies averaged a depth of 3.41 and 9.39 leaves, yielding hierarchies that extended well beyond simple flat lists. Most structures were shallow (2--5 levels), although a few reached greater depth. Broad domains produced larger taxonomies (e.g., \textit{image generation}, depth 11 with 438 leaves; \textit{language translation}, depth 10 with 241 leaves), whereas narrower domains resulted in more compact structures (e.g., \textit{translation assistant}, depth 4 with 12 leaves). Notably, no empty taxonomies were produced, underscoring the robustness of the pipeline.  

\begin{tcolorbox}[takeaway]
\textbf{Takeaway 4.} The framework generates interpretable multi-level taxonomies that adapt to domain scope, supporting both high-level overviews and fine-grained feature analysis, while ensuring robust outputs across applications.
\end{tcolorbox}

\begin{table}[t]
\centering
\caption{Qualitative analysis of top semantically ranked clusters per app.}
\begin{tabular}{p{1.5cm}|c|p{2.0cm}|c|p{4.1cm}|p{2.6cm}}
\toprule
\textbf{App} & \textbf{\#C} & \textbf{Cluster} & \textbf{\#F} & \textbf{Features} & \textbf{Match} \\
\midrule
ChatGPT & 2168 & Advice & 12  & \textit{advise, advice, guiding, guidance, ...} & Guidance and recommendations \\
\midrule
Claude & 517 & AI-assisted writing\newline support & 8  & \textit{computer languages, software development, programming, code generation, coding, ...} & Programming and code assistance \\
\midrule
DeepSeek & 676 & AI interface & 18  & \textit{ai apps, ai interface, ai chat, coding, multi languages, learning, intelligence, ...} & AI interaction and learning \\
\midrule
Google Gemini & 1571 & Activities\newline meetings\newline working & 13 & \textit{trips, gym, meet, office, working, tasks, activities, meetings, ...} & Task and activity management \\
\midrule
Le Chat & 144 & Bot & 5 & \textit{prompt, confirmation, bot, chat, message} & Chat interaction \\
\midrule
Microsoft Copilot & 1115 & AI\newline applications& 12  & \textit{ai support, ai companion, ai search, ai platform, ai chat, ai assistant, ...} & AI tool variety \\
\midrule
Perplexity & 936 & Analysis and visualization & 8  & \textit{chart, data, statistics, analysis, data analysis, ...} & Data analysis and charts \\
\bottomrule
\end{tabular}
\label{tab:cluster_validation}
\end{table}

For qualitative validation, we first computed semantic coherence scores for all clusters by measuring cosine similarity between embeddings of LLM-generated cluster labels and their member features. We then compared the top-ranked clusters against the official Google Play App Store application descriptions. Table~\ref{tab:cluster_validation} presents one representative cluster per app, showing: (1) the mean number of clusters generated when analyzing the full dataset with Hybrid feature extraction (combining T-Frex and TransFeatEx) and the balanced clustering strategy (2) the best-scoring cluster name, (3) a representative (non-exhaustive) list of features, and (4) the corresponding match from the official documentation. Due to space constraints, we report only one cluster per app.%, while the data containing the top 20 clusters is available in the replication package.

\begin{figure}[hbtp]
    \centering
    \includegraphics[width=0.8\linewidth]{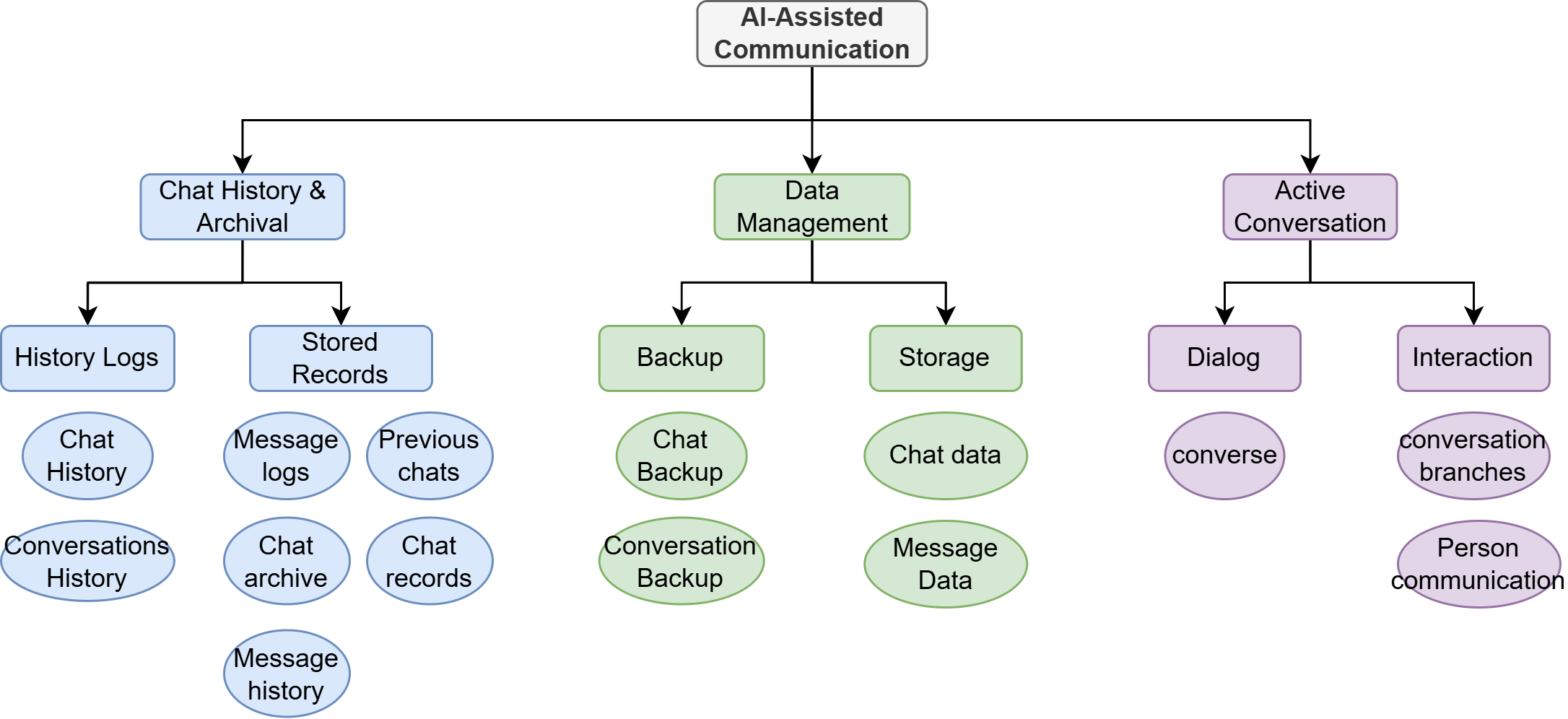}
    \caption{Extracted generated taxonomy for a high semantic scoring cluster.}
    \label{fig:example_taxonomies}
\end{figure}

Figure~\ref{fig:example_taxonomies} illustrates hierarchical structures for one of the high-scoring clusters. The \textit{AI-Assisted Communication} taxonomy organizes AI interaction and capability features, spanning from user-facing applications and communication tools to the underlying programming languages and intelligence technologies. The intermediate nodes represent subcategories automatically tagged by the LLM to capture thematic groupings, and the leaf nodes correspond to concrete features extracted from reviews.

While the resulting taxonomies in this example generally align well with intuitive semantic relations, minor inconsistencies appear. For example, \textit{"message history"} might better fit under history logs, and the boundary between \textit{"dialog"} and \textit{"interaction"} is somewhat blurred. These small deviations reflect the inherent limitations of hierarchical clustering, but overall the taxonomy remains coherent and meaningful from a human--computer interaction perspective.

\begin{tcolorbox}[takeaway]
\textbf{Takeaway 5.} Structured feature taxonomies can be aligned with official mobile app documentation, demonstrating the applicability of \textit{FeClustRE} in real-world scenarios and enabling better interpretation by RE practitioners.
\end{tcolorbox}

Due to space constraints, the complete taxonomies and results are available in our replication package. %\footnote{Available at: [URL]}.
\section{Threats to Validity}

Concerning construct and design validity, the framework's performance depends on the quality of initial feature extraction, with TransFeatEx and T-FREX chosen as representative syntactic- and LLM-based methods. 
The hybrid design may compound individual method errors, though our results suggest that the benefits outweigh these risks. 
Larger datasets increase computational cost but also yield richer and more interpretable taxonomies, justifying this trade-off. 
For embeddings, we employed Sentence-BERT and T5-Sentence, and used silhouette and Davies--Bouldin indices for clustering evaluation; while standard, these do not capture all aspects of interpretability. 
Parameter sensitivity in hierarchical clustering (cut-off $\tau$, sibling $s_t$) and few-shot prompt design are additional sources of variation. 
The few-shot strategy, while effective with 3--5 examples per domain, may benefit from more systematic optimization across categories.

Concerning external validity, annotation biases in expert and crowdsourced datasets for RQ\textsubscript{1} may affect evaluation outcomes, as annotation processes introduce artifacts. 
Manual cluster inspection by the authors also introduces evaluator bias, partially mitigated through the alignment strategy with official documentation; independent third-party evaluation would further strengthen objectivity. 
Our focus on generative LLM-based chatbot apps limits generalizability, as their technical terminology and evolving feature sets may differ from other domains.

Finally, concerning conclusion validity, our evaluation design introduces further limitations. 
The choice of correctness scores ($P$, $R$, $F_\beta$) and clustering metrics reflects our construct decisions and may emphasize certain aspects of performance. 
The 2K vs.~50K sampling strategies allowed us to examine data quantity effects, but results could vary under other schemes. 
The dataset's temporal scope (July 2025) and relatively small number of apps (seven) also restrict generalizability. 
Nevertheless, with 158,207 reviews, the dataset provides substantial evidence to support the robustness of our evaluation.

\section{Conclusions}

In this work, we proposed a hybrid method that combines syntactic feature extraction with LLM-based analysis and hierarchical clustering to improve correctness and interpretability in app review feature mining. Our contribution includes: (1) a hybrid framework combining syntactic precision with semantic understanding through automated clustering and LLM tagging, (2) an auto-tuning mechanism for optimal cluster configuration with a comprehensive evaluation methodology, and (3) an open-source implementation with graph-based storage.

The experimental validation confirms that our hybrid approach achieves the best balance of performance, with consistently higher recall and $F_\beta$ scores compared to syntactic or LLM-only baselines. The auto-tuning component further avoids manual parameter setting and adapts across app categories and dataset sizes, producing coherent structures from both small and large-scale datasets. Moreover, the hybrid configuration generated richer and more fine-grained taxonomies while maintaining stable clustering quality.

Our evaluation of LLM-based semantic tagging showed high distinctiveness and coherence of taxonomy labels, strengthened by the merging mechanism that reduced redundancy without compromising semantic integrity. Qualitative validation against Google Play documentation demonstrated that cluster labels align with established feature descriptions, ensuring reliability beyond subjective interpretation. The resulting taxonomies not only capture semantically related features but also exhibit multi-level structures, organizing concepts from general categories down to specific functionalities. In practice, these taxonomies enable practitioners to prioritize features, detect trends, and compare applications, supporting systematic market analysis and release impact assessment.

FeClustRE thus provides value for requirements engineering tasks through interpretable feature taxonomies and interactive usage. Current limitations include dependency on initial feature extraction quality, computational complexity for large-scale datasets, and sensitivity to clustering parameters. Future work will explore optimal configurations across app categories, domain-dependent thresholds for ranking clusters, and extending applicability to complex RE tasks such as competition and market analysis, further improving the utility and robustness of app review feature extraction pipelines.

\section*{Data Availability Statement}

Full source code, datasets, clusters, prompts, and evaluation reports are available in a replication package \href{https://doi.org/10.5281/zenodo.17372435}{https://doi.org/10.5281/zenodo.17372435}. Additionally, a  GitHub repository is also available \href{https://github.com/nlp4se/FeClustRE}{https://github.com/nlp4se/FeClustRE}.

% --- SciKGTeX Open Science Annotations ---
%\contribution{replication package}{https://doi.org/10.5281/zenodo.17372435}
%\contribution{code repository}{https://github.com/nlp4se/FeClustRE}
%\contribution{dataset}{https://doi.org/10.5281/zenodo.17372435}
% --- End SciKGTeX annotations ---

%\input{content/B_acknowledgments}

\bibliographystyle{splncs04}
\bibliography{ref}

\end{document}